\documentclass{aastex631}

\usepackage{graphicx}
\usepackage{amsmath}
\usepackage{longtable}
\usepackage{color}
\usepackage{enumerate}
\usepackage{graphicx,epstopdf}
\usepackage{cases}
\usepackage{multirow}
\usepackage{tikz,xcolor,hyperref}
\usepackage{lipsum}

\shorttitle{A WD-MS binary system}
\shortauthors{Zheng et al.}

\begin{document}

\title{A White Dwarf-Main Sequence Binary Unveiled by Time-Domain Observations from LAMOST and TESS}

\correspondingauthor{Wei-Min Gu}
\email{guwm@xmu.edu.cn}

\author[0000-0002-5630-7859]{Ling-Lin Zheng}
\affiliation{Department of Astronomy, Xiamen University,Xiamen, Fujian 361005, P. R. China}
\author[0000-0003-3137-1851]{Wei-Min Gu}
\affiliation{Department of Astronomy, Xiamen University,Xiamen, Fujian 361005, P. R. China}
\author[0000-0002-0771-2153]{Mouyuan Sun}
\affiliation{Department of Astronomy, Xiamen University,Xiamen, Fujian 361005, P. R. China}
\author[0000-0002-2419-6875]{Zhixiang Zhang}
\affiliation{Department of Astronomy, Xiamen University,Xiamen, Fujian 361005, P. R. China}
\author[0000-0002-5839-6744]{Tuan Yi}
\affiliation{Department of Astronomy, Xiamen University,Xiamen, Fujian 361005, P. R. China}
\author[0000-0001-7349-4695]{Jianfeng Wu}
\affiliation{Department of Astronomy, Xiamen University,Xiamen, Fujian 361005, P. R. China}
\author[0000-0003-4874-0369]{Junfeng Wang}
\affiliation{Department of Astronomy, Xiamen University,Xiamen, Fujian 361005, P. R. China}
\author[0000-0003-2896-7750]{Jin-Bo Fu}
\affiliation{Department of Astronomy, Xiamen University,Xiamen, Fujian 361005, P. R. China}
\author[0000-0002-7135-6632]{Sen-Yu Qi}
\affiliation{Department of Astronomy, Xiamen University,Xiamen, Fujian 361005, P. R. China}
\author[0000-0002-6039-8212]{Fan Yang}
\affiliation{Department of Astronomy, Beijing Normal University, Beijing 100875, China}
\author[0000-0003-3116-5038]{Song Wang}
\affiliation{National Astronomical Observatories, Chinese Academy of Sciences, Beijing 100101, China}
\author[0000-0003-3603-1901]{Liang Wang}
\affiliation{Nanjing Institute of Astronomical Optics \& Technology, Chinese Academy of Sciences, Nanjing 210042, China}
\affiliation{CAS Key Laboratory of Astronomical Optics \& Technology, Nanjing Institute of Astronomical Optics \& Technology, Chinese Academy of Sciences, Nanjing 210042, China}
\author{Zhongrui Bai}
\affiliation{National Astronomical Observatories, Chinese Academy of Sciences, Beijing 100101, China}
\author[0000-0002-6617-5300]{Haotong Zhang}
\affiliation{National Astronomical Observatories, Chinese Academy of Sciences, Beijing 100101, China}
\author[0000-0002-6647-3957]{Chun-Qian Li}
\affiliation{National Astronomical Observatories, Chinese Academy of Sciences, Beijing 100101, China}
\author[0000-0002-0349-7839]{Jian-Rong Shi}
\affiliation{National Astronomical Observatories, Chinese Academy of Sciences, Beijing 100101, China}
\author[0000-0002-7660-9803]{Weikai Zong}
\affiliation{Department of Astronomy, Beijing Normal University, Beijing 100875, China}
\author[0000-0002-0349-7839]{Yu Bai}
\affiliation{National Astronomical Observatories, Chinese Academy of Sciences, Beijing 100101, China}
\author[0000-0002-2874-2706]{Jifeng Liu}
\affiliation{National Astronomical Observatories, Chinese Academy of Sciences, Beijing 100101, China}
\affiliation{College of Astronomy and Space Science, University of Chinese Academy of Sciences, Beijing 100049, China}

\begin{abstract}
We report a single-lined white dwarf-main sequence binary system, LAMOST
J172900.17+652952.8, which is discovered by LAMOST’s medium resolution 
time-domain surveys. The radial velocity semi-amplitude and orbital period 
of the optical visible star are measured by using the Palomar 200-inch 
telescope follow-up observations and the light curves from \textit{TESS}. 
Thus the mass function of the invisible candidate white dwarf is derived, 
$f(M_{\rm{2}}) = 0.120\,\pm\,0.003\,M_{\odot}$.
The mass of the visible star is measured based on the spectral energy
distribution fitting, $M_{\mathrm{1}}$ = $0.81^{+0.07}_{-0.06}\,M_{\odot}$.
Hence, the mass of its invisible companion is 
$M_{\rm{2}}\,\gtrsim\,0.63\,M_{\odot}$. The companion ought to be a 
compact object rather than a main-sequence star owing to the mass ratio
$q = M_{\rm{2}} / M_{\rm 1} \gtrsim 0.78$ and the single-lined spectra. 
The compact object is likely to be a white dwarf except for small inclination
angle, $i\,\lesssim\,40^{\circ}$. By using the GALEX NUV flux, the 
effective temperature of the white dwarf candidate is constrained as 
$T_{\rm eff}^{\rm WD}\,\lesssim\,12000-13500$ K. It is difficult to detect 
white dwarfs which are outshone by their bright companions via single-epoch 
optical spectroscopic surveys. Therefore, the optical time-domain surveys can 
play an important role in unveiling invisible white dwarfs and other compact 
objects in binaries.
\end{abstract}


\section{Introduction}\label{sec1}
A white dwarf-main sequence (WD-MS) binary is a system composed of a WD
and a main-sequence star. The study of WDs can facilitate us
to understand the physical properties of binaries and the evolution of the
Milky Way \citep[e.g.,][]{Ren2018}. If the separation of two stars is far
enough ($\gtrsim$ 10 AU) that they can evolve separately, the more massive
one ($\lesssim$ 10 $M_{\odot}$) will reach the end of its life as a WD
earlier than its less massive companion \citep[e.g.,][]{Rebassa-Mansergas2021}.
The stellar evolution model predicts that more than 97\% of the stars with
masses $\lesssim$ 10 $M_{\odot}$ in our Galaxy eventually form as WDs
\citep[e.g.,][]{Garcia1997,Fontaine2001,Taani2017,Kepler2019}.

  The unique characteristics of WDs, such as the blue colors and relatively
large proper motions, have traditionally been used to identify WDs
\citep[e.g.,][]{Raddi2017}. Over the past decades, optical telescopes with
high-resolution, high-precision, and wide-field, have led to a significant
increase in the sample size of WD-MS binaries  \citep[e.g.,][]{Silvestri2006,Heller2009,Liu2012,Li2014,Cojocaru2017,Bar2017,Ren2018}. 
Most of these binaries are identified by the prominent lines of WDs shown in
the spectra, e.g., Balmer lines in the blue spectra for the hot WDs 
\citep[the effective temperature $\gtrsim$ 15000 K,][]{Wesselius1978}. 
However, due to the faint nature of WDs, the search methods relying on the 
spectral components of WDs are often limited to nearby sources
\citep[e.g.,][]{Bar2017,Kilic2020,Rebassa-Mansergas2021}.
The Sloan Digital Sky Survey (SDSS) found a large WD sample
including hot WDs \citep[e.g.,][]{Rebassa-Mansergas2016}. Such hot WDs are 
sometimes indistinguishable from quasars in color space. Hence, a large 
fraction of the identified and known WD-MS binaries in SDSS are composed of 
relatively cool WDs with late-type companions \citep{Rebassa2012,Ren2018,Rebassa-Mansergas2021}. Motivated by complementing 
the sample of WDs with bright companions (e.g., F, G, and K type star), some 
research attempt to select the sources based on the conditions of UV excess 
and the stellar temperature 
\citep[e.g.,][]{Maxted2009,Parsons2016,Rebassa-Mansergas2017,Hernandez2021}
or the multi-band photometry \citep[e.g.,][]{Maxted2009}.

  The superiority of optical time-domain surveys, including spectroscopy and
photometry, allow us to make continuous tracking of the observed targets over
time. We can derive the semi-amplitude $K$ of radial velocity $V_{\rm r}$ 
and photometric variation period $T_{\rm ph}$ of a binary system from 
time-domain surveys. The time-domain method has the potential to unveil 
distant faint WDs, and enlarge the sample of WDs with bright companions 
(e.g., A, F, G, and K type star) even without the necessary characteristics 
(e.g., UV excess or blue color).

   A multi-object time-domain spectroscopic survey would be vital for
constraining $K$ (or even $T_{\rm ph}$) for a tremendous amount of binaries.
LAMOST (Large Sky Area Multi-Object fiber Spectroscopic
Telescope)\footnote{\url{http://www.lamost.org/public/?locale=en}} is an
optical ground-based telescope with wide-field of view ($\sim\,5^{\circ}$)
and a large number of fibers
$\sim\,4000$ \citep{Cui2012,Zhao2012,Liu2020,Zong2018,Zong2020}. Not only
does it provide us with a huge number of stellar spectra, but also it allows
for multiple observations of the same target. Notably, the LAMOST
medium-resolution survey (hereafter MRS, whose spectral resolution
$R\sim$ 7500, G $\lesssim$ 15 mag) will complete about 60 high-quality
spectroscopic observations for each of 200,000 sources (G $\sim$ 14 mag)
within five years \citep[2018-2023;][]{Liu2020}. This time-domain survey will
be of great help to our search for compact binaries, including binaries with
WDs, neutron stars (NS), or black holes \citep[e.g.,][]{Yi2019}. Up to the
Data Release 8 (DR8), LAMOST has released more than 10 million stellar
spectra\footnote{\url{http://www.lamost.org/dr8/v1.1/}}, and it has some 
fruitful results in the search for compact binaries or candidates \citep[e.g.,][]{Gu2019,Liu2019,Zheng2019,Wiktorowicz2020,Yang2021,Zhang2022}.

  For sources with multiple spectroscopic observations from LAMOST,
\textit{TESS} (Transiting Exoplanet Survey
Satellite)\footnote{\url{https://tess.mit.edu/}}
survey can serve as a perfect complement in photometry. \textit{TESS} is an
all-sky survey spacecraft to collect the photometric data for exoplanets
transiting bright and nearby stars \citep{Ricker2015}, and its detection limit
is about 15 \textit{TESS} magnitude. \textit{TESS} performs both 2-minute and
30-minute cadence observations \citep{Stassun2018}. Hence, the \textit{TESS}
observations are vital for probing short-period binaries.

   The joint effort of LAMOST and \textit{TESS} has great potential for
hunting for short-period binaries that host compact objects. \citet{Mu2022}
presents a sample of compact object candidates with K/M-dwarf companions by
using radial velocities from LAMOST and the photometric period from
\textit{TESS}. Here we report the discovery of a new object LAMOST,
J172900.17+652952.8 (hereafter J1729+6529), which is a WD-MS binary system
with a bright K-type companion discovered by radial velocities and periodic
light curves. The object has Palomar 200-inch telescope (P200) follow-up
observations, as well as 20 sectors of \textit{TESS} observations 
(each sector is 27.4 days long).

  We organize the paper as follows. The observations and data reduction are
presented in Section~\ref{sec2}. Results are described in Section~\ref{sec3}.
In Section~\ref{sec4}, we discuss the implications of our results.

\begin{figure*}[htb]
\plotone{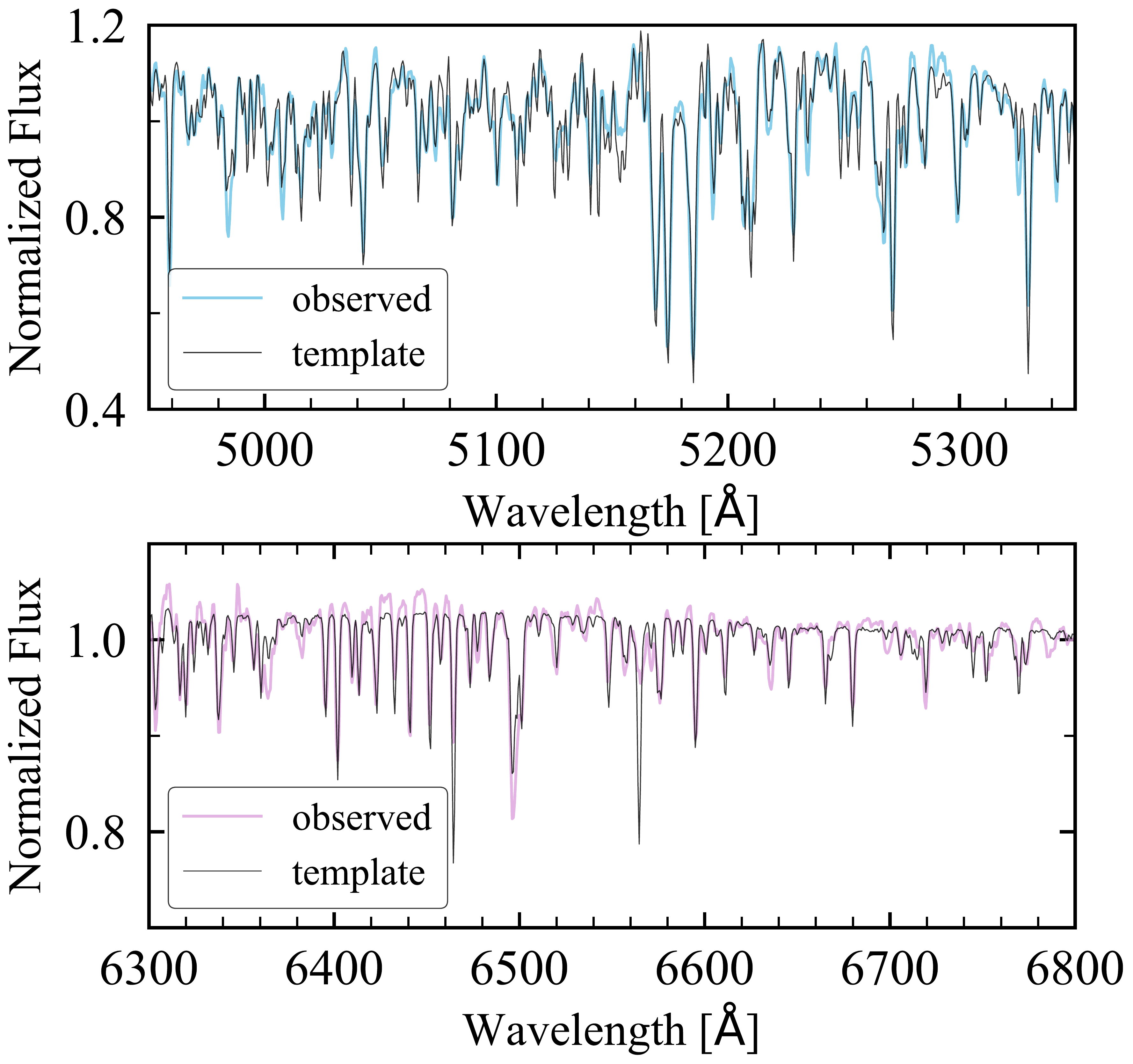}
\caption{The LAMOST coadd spectrum (blue and pink curves) of J1729+6529 and
the best-matching template (black curves). The upper and lower panels are for
the blue and red arms, respectively.
\label{F1}}
\end{figure*}

\section{Observations and data reduction}\label{sec2}
\subsection{The discovery of J1729+6529}

   J1729+6529 is a single-lined object selected from the medium-resolution
survey of LAMOST DR7, whose equatorial coordinate is
R.A. = 262.25073$^{\circ}$ and Decl. = +65.49800$^{\circ}$ in J2000. This
source draws our attention for the rapid variation of radial velocity, i.e.,
$\left| \Delta V_{\rm r} \right| \textgreater 100$ km\,s$^{-1}$ within 141
minutes. We judge the spectra of the source are single-lined by using the
cross correlation functions
\citep[CCFs; refer][]{Merle2017,Li2021}.

  We collect the photometric data of J1729+6529 from different sky surveys,
including \textit{TESS}, ASAS-SN, and ZTF. We search the MIKULSKI ARCHIVE
FOR SPACE TELESCOPES (MAST)
\footnote{\url{https://mast.stsci.edu/portal/Mashup/Clients/Mast/Portal.html}}
with a maximum searching radius of 0.5$\arcsec$, and found that our target
has 20 sectors of \textit{TESS} observations. We derive the processed
light-curve data of the sectors for 14-19, 21-22, and 24-26 directly from the
MAST. Then fold the photometric data according to the period revealed by the
Lomb-Scargle algorithm, which is a method by means of the Fourier transform
to extract periodic signals in the non-uniform time domain
\citep{Lomb1976,Scargle1981}. We find that all folded light curves show
periodic signals.

  Due to only several LAMOST $V_{\rm r}$ measurements for J1729+6529 
on the same night, the semi-amplitude $K$ cannot be constrained reliably. 
Hence, we implement the follow-up observations of J1729+6529 by using DBSP 
(Double Spectrograph) of P200 telescope \citep{Oke1982} on June 12 and 24, 
2020. The seeing is about 1.3$\arcsec$ during the observation. The spectral 
coverage is [4047.9 \AA, 5594.9 \AA] with a resolving power of 3415 for the 
blue-side, and [5788.4 \AA, 7089.5 \AA] with a resolving power of 4524 for 
the red-side.

  We follow the standard procedures and use
IRAF\footnote{\url{https://github.com/joequant/iraf}} to extract the 1-D
spectra. Then we reject the cosmic rays of the observed spectra by using 
the median filter, and all spectra are normalized by the Gaussian filter 
method. We use The\_Payne \citep{Ting2019} to interpolate the BT-Settl 
spectral grid, which is a spectral interpolate tool based on neural networks. 
To obtain the best matching template, we minimize the $\chi^2$ of the 
template model and the observed spectrum. The $\chi^2$ is expressed as
\begin{equation} 
\chi^2 = \sum\limits_{i}\,\left(\frac{y_{\rm i}-y_{\rm model}}{\sigma_{i}}\right)^2,
\label{E1}
\end{equation}
where $y_{\rm i}$ is the spectral flux of the i-th point, and $y_{\rm model}$
is the spectral flux of the model. The best matching spectral template (see
the black curves of Figure~\ref{F1}), which corresponds to the smallest value
of the $\chi^2$, is selected by interpolating the grid from the BT-Settl 
model. Next, we use the cross-correlation function
\citep[CCF;][]{Tonry1979,Gaskell1987,Zverko2007} in the
\textit{spectool} package\footnote{\url{https://gitee.com/zzxihep/spectool}}
to measure $V_{\rm r}$ of LAMOST and P200 by cross-matching the template and
the observed spectra. The blue and pink curves in Figure~\ref{F1} are selected
from LAMOST, shown as an example. $V_{\rm r}$ can be calculated by
$V_{\mathrm{r}}$ = $c\,\Delta \lambda / \lambda$ = $c\,\Delta \rm{ln}\lambda$ = $c\,\Delta$u, 
where $c$ is the speed of light, $\lambda$ is the wavelength. The CCF is 
defined by the following equation:
\begin{equation}
\rm CCF(\Delta \rm{u})
= \int \frac{[\it{f}(\rm{u})-\overline{\it{f}}] [\it{T}(\rm{u}-\Delta u)-\overline{\it{T}}]} {\sigma\it{_f}\,\sigma\it{_T}}\,d\rm{u},
\label{E2}
\end{equation}
where $f(\rm{u})$ and $T(\rm{u})$ are the observed and template spectra 
in $\log$ wave space; $\sigma\it{_f}$ and $\sigma\it{_T}$ are the standard 
deviations of the observed spectrum and the template, respectively. The
sampling interval of the $V_{\rm r}$ is set as 1 km s$^{-1}$. Since the
neon bulb (for the HeNeAr lamp) was broken during the P200 observation, we
only take the blue-side of the P200 spectra to measure the corresponding
$V_{\rm r}$. The used wavelength windows in the measured process are
[5200 \AA, 5350 \AA] and [5200 \AA, 5500 \AA] for the spectra of LAMOST and 
P200, respectively. The uncertainties of $V_{\rm r}$ are measured 
by using the ``flux randomization/random subset sampling (FR/RSS)" method  
\citep{Peterson1998}, which is a Monte Carlo approach. For each Monte Carlo 
simulation, we add a random flux to each pixel based on the 1$\sigma$ 
flux uncertainties of that pixel (i.e., the ``FR" step), and randomly 
removes data points in the spectrum (i.e., the ``RSS" step). We run the 
Monte Carlo simulation 1000 times to obtain the mock $V_{\mathrm{r, m}}$. 
The uncertainty of $V_{\mathrm{r}}$ is the standard deviation of the 
distribution of $V_{\mathrm{r, m}}$. $V_{\rm r}$ measurements are shown in
Table~\ref{Vr:measures}.

\subsection{The determination of the orbital period and the radial-velocity
semi-amplitude}\label{sec2.2}

  The photometric periods from \textit{TESS}, ASAS-SN, and ZTF are all
different. These three periods measured by the Lomb-Scargle algorithm are 
0.6 days, 3 days and 1.2 days, respectively. We derive the true orbital 
period of J1729+6529 is 0.6 days by fitting the radial velocities, the 
details of which are discussed below.

\begin{figure*}[htb]
\plotone{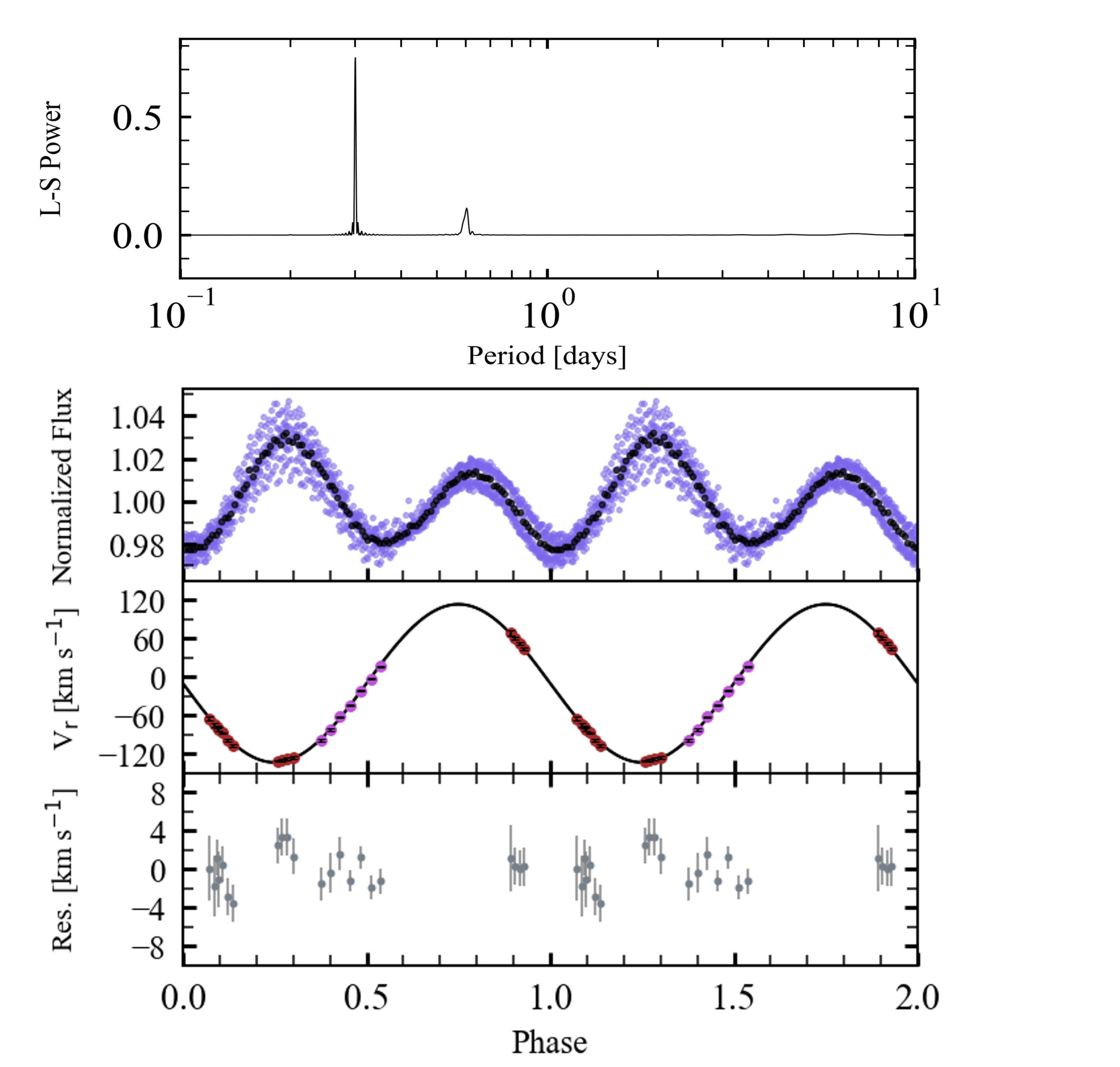}
\caption{The first panel (from top to bottom) shows the Lomb-Scargle 
power peak. The second panel displays the \textit{TESS} light curve 
folded with the orbital period of 0.600303 days. The black dots are 
binned over ten violet dots. The third panel is the radial velocities 
folded by the period of 0.600303 days. The black curve is the 
best-fitting radial-velocity curve with a semi-amplitude of 
$K = 124.4^{+1.0}_{-1.1}$ km s$^{-1}$. The pink dots represent the data 
collected from LAMOST, and the dark red dots are from P200. The errors of 
the photometric data are too small to be shown in the light curve, and the 
errors of the radial velocities are shown with the short bars over the dots 
(see the third panel). The residuals in the fourth panel are obtained 
by subtracting the observed radial velocities from the best fit model 
(Equation~(\ref{E3})). It should be noted that the eccentricity is fixed to 
zero when fitting the radial velocities.
\label{F2}}
\end{figure*}

  The radial velocity fit for the visible star is conducted by using
\textit{The Joker}\footnote{\url{https://github.com/adrn/thejoker}}
\citep{Price2017}. The formula used for the fit is expressed as follows:
\begin{equation}
V_{\mathrm{r}}(\it t) = \it v_{\rm 0} + \it K [\cos(\it f + \omega) + \it e\cos(\omega)],
\label{E3}
\end{equation}
where $v_{\rm 0}$, $K$, $f$, $\omega$, and $e$ represent the systemic
velocity, the velocity semi-amplitude, the true anomaly, the argument of
periastron, and eccentricity, respectively. The eccentricity is fixed 
to zero, and the calibration offset between instruments of LAMOST and P200
is set as a free parameter in the fitting process. The prior of the offset 
is assumed to follow the Gaussian distribution, whose mean value and standard 
deviation are $0\ \mathrm{km\ s^{-1}}$ and $5\ \mathrm{km\ s^{-1}}$, 
respectively\footnote{\url{https://thejoker.readthedocs.io/en/latest/examples/5-Calibration-offsets.html}}.
Due to our incomplete understanding of the radial-velocity uncertainties 
and the intrinsic radial-velocity scatter, there may exist a jitter in the 
radial-velocity curve \citep{Price2017}. We consider this jitter $s$ in the 
fitting, and the derived value is given in Table~\ref{tbl:parameter}. 
The fitting results (see Table~\ref{tbl:parameter}) suggest that models with 
periods of 3 days and 1.2 days cannot explain the observed $V_{\mathrm{r}}$. 
In contrast, the model with the prior period from \textit{TESS} shows 
a perfect fit to the radial velocities (see the third panel of 
Figure~\ref{F2}, from top to bottom). Thus we argue that the orbital period 
with its 1$\sigma$ uncertainty is $0.600303(6)$ days (given by 
\textit{The Joker}), which is consistent with the \textit{TESS}'s photometric 
period 0.600149 days (the highest peak in the first panel of 
Figure~\ref{F2}). The third panel is the radial-velocity curve folded 
by the orbital period $P_{\rm orb}$, and the corresponding semi-amplitude of 
the best-fitting radial-velocity curve (black curve) is 
$K = 124.4^{+1.0}_{-1.1}$ km s$^{-1}$. The fourth panel gives the 
residuals obtained by subtracting the theoretical $V_{\rm r}$ 
(Equation~(\ref{E3})) from that of the observed. The uncertainties of the 
residuals are calculated by $\sqrt{(V_{\rm r,err})^{2} + s^{2}}$, where 
$V_{\rm r,err}$ is the uncertainty of the $V_{\rm r}$.

\begin{table}[htb]
\begin{center}
  \caption{Observation log of the $V_{\rm r}$}
  \label{Vr:measures}
  \begin{tabular}{ccccc}
  \hline
  \hline
Instrument  & Median UTC & Phase & Exposure time & $V_{\rm r}$ \\  
 & (yyyy-mm-dd hh:mm:ss) &  &  (s)  &  (km s$^{-1}$)  \\
  \hline
	\multirow{7}{*}{LAMOST}	&	2019-05-27 16:40:00	& 	0.37 	& 	1200	&	-98.7	$\pm$	1.7 	 \\
&	2019-05-27 17:04:00	&	0.40 	&	1200	&	-81.9	$\pm$	2.1 	 \\
&	2019-05-27 17:27:00	&	0.43 	&	1200	&	-62.1	$\pm$	1.7 	 \\
&	2019-05-27 17:51:00	&	0.46 	&	1200	&	-44.4	$\pm$	1.1 	 \\
&	2019-05-27 18:14:00	&	0.48 	&	1200	&	-21.7	$\pm$	1.2 	 \\
&	2019-05-27 18:38:00	&	0.51 	&	1200	&	-2.9	$\pm$	1.2 	 \\
&	2019-05-27 19:00:00	&	0.54 	&	1053	&	16.9 	$\pm$	1.3 	 \\
\hline
	\multirow{15}{*}{P200}	&	2020-06-12 04:49:25	&	0.89 	&	60	&	62.3 	$\pm$	3.4 	 \\
&	2020-06-12 04:59:30	&	0.90 	&	600	&	54.3 	$\pm$	1.9 	 \\
&	2020-06-12 05:10:11	&	0.92 	&	600	&	45.3 	$\pm$	1.9 	 \\
&	2020-06-12 05:20:33	&	0.93 	&	600	&	37.6 	$\pm$	2.0 	 \\
&	2020-06-12 07:44:33	&	0.09 	&	720	&	-84.6	$\pm$	1.8 	 \\
&	2020-06-12 07:56:55	&	0.11 	&	720	&	-93.6	$\pm$	1.9 	 \\
&	2020-06-12 08:09:17	&	0.12 	&	720	&	-105.6	$\pm$	1.9 	 \\
&	2020-06-12 08:21:38	&	0.14 	&	720	&	-113.4	$\pm$	1.9 	 \\
&	2020-06-12 10:04:28	&	0.26 	&	720	&	-137.3	$\pm$	1.8 	 \\
&	2020-06-12 10:16:49	&	0.27 	&	720	&	-135.5	$\pm$	1.9 	 \\
&	2020-06-12 10:29:11	&	0.28 	&	720	&	-133.5	$\pm$	1.9 	 \\
&	2020-06-12 10:41:33	&	0.30 	&	720	&	-132.5	$\pm$	1.8 	 \\
&	2020-06-24 07:34:46	&	0.07 	&	600	&	-71.0 	$\pm$	3.4 	 \\
&	2020-06-24 07:45:08	&	0.08 	&	600	&	-80.8	$\pm$	3.2 	 \\
&	2020-06-24 07:55:29	&	0.10 	&	600	&	-87.8	$\pm$	2.9 	 \\
  \hline
  \end{tabular}
  \\
  \textbf{Notes.}
  Instrument: the telescopes used for the observations of J1729+6529.
  Median UTC: the intermediate time between the start and the end of the 
  observation. \\ 
  Phase: the orbital phase. \\
  Exposure time: duration of the observation. \\
  $V_{\rm r}$: the radial velocity measured by the CCF. \\
\end{center}
\end{table}

\begin{table}[htb]
\begin{center}
  \caption{\textbf{\textit{The Joker} fitting parameters}}
  \label{tbl:parameter}
  \begin{tabular}{cc}
  \hline
  \hline parameter & value and uncertainty \\ \hline
    (1) & (2) \\ \hline
  $P_{\rm orb}$ [days] 	&	$0.600303(6)$ \\
  $K$ [km s$^{-1}$]	&	 $124.4^{+1.0}_{-1.1}$ \\
  $v_{\rm 0}$ [km s$^{-1}$]	&	 $-9.6^{+4.0}_{-3.9}$	\\
  $s$ [km s$^{-1}$]	      &       $0.1^{+0.3}_{-0.1}$   \\
  $d\rm{v0}$ [km s$^{-1}$]   &   $-4.2^{+3.6}_{-3.5}$\\
  $\omega$ [rad]	 &	 $3.1 \pm 2.0$ \\
  $f(\it M_{\rm 2})$  [$M_{\odot}$]	&  $0.120 \pm 0.003$  \\ 
  $e$   	&	 	fixed to 0      \\
  $\chi^2$  &   19.4  \\
  $N_{\it V_{\rm r}}$  &   22 \\
  \hline
  \end{tabular}
  \\
  \textbf{Notes.}
  $P_{\rm orb}$: the orbital period.\\
  $K$: the radial velocity semi-amplitude of the visible star.\\
  $v_{\rm 0}$: the systemic velocity.\\
  $s$: the jitter of the radial-velocity curve.\\
  $d\rm{v0}$: the calibration offset between LAMOST and P200.\\
  $\omega$: the argument of periastron.\\
  $f(M_{\rm 2})$: the mass function. \\
  $e$: the orbit eccentricity.\\
  $\chi^2$: the Chi-square value of the radial-velocity fitting.\\
  $N_{\it V_{\rm r}}$: the number of radial velocities.\\
\end{center}
\end{table}

\begin{figure*}[htb]
\plotone{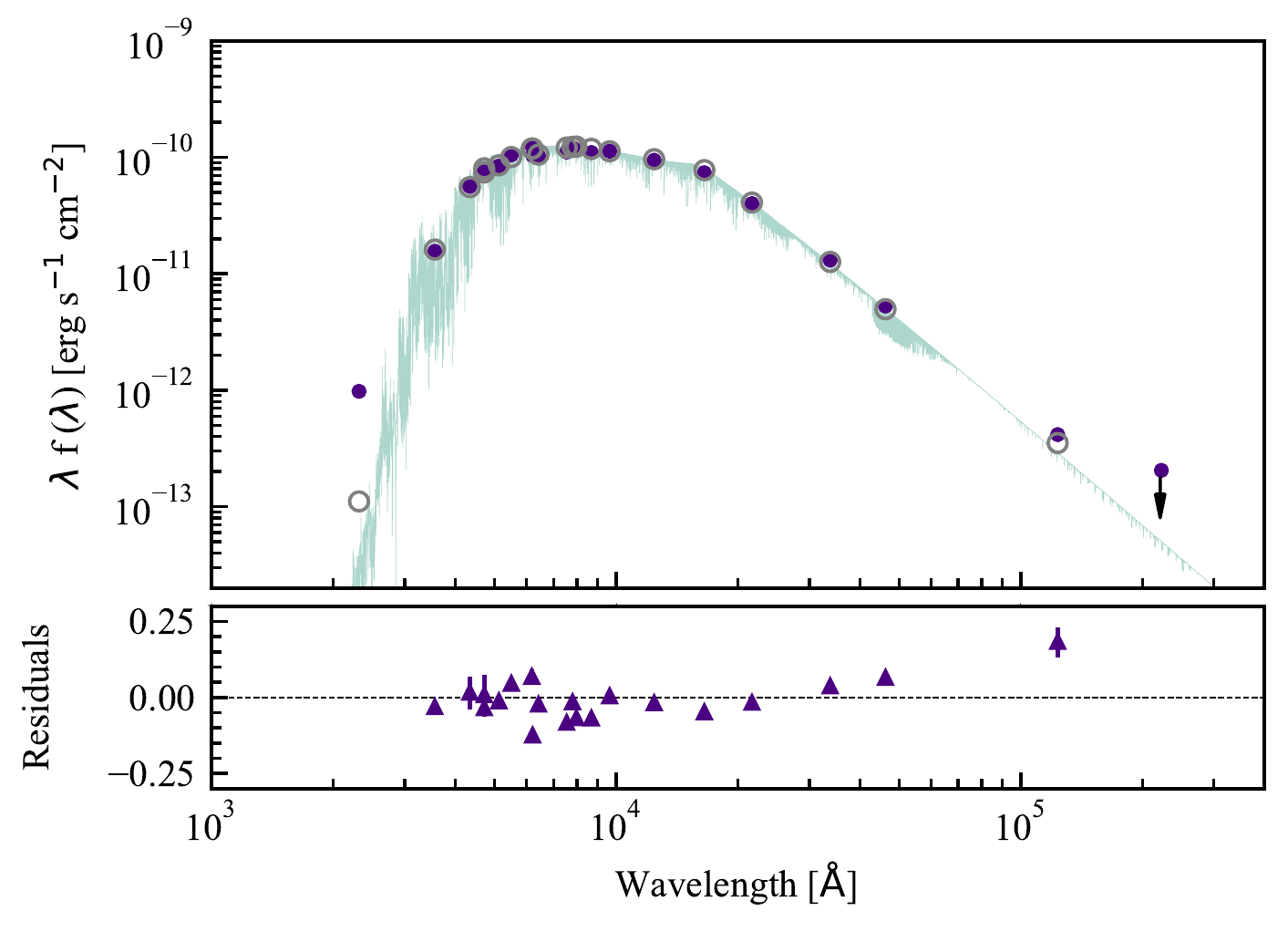}
\caption{The spectral energy distribution of J1729+6529. The dark violet
dots are the photometric data collected from \textit{GALEX}, SDSS, APASS
g and r bands, APASS GROUND JOHNSON B and V, Pan-STARRS,
\textit{Gaia}, \textit{TESS}, 2MASS, and ALLWISE. The gray circles are the
synthetic fluxes from the best-fitting theoretical spectral template. The
black arrow indicates an upper limit of the WISE W4 band. The 
green curve is the best-fitting model. The lower panel shows 
the residuals calculated by the difference of the photometric and the 
synthetic fluxes divided by the synthetic fluxes. It should be noted that 
NUV, W3, and W4 were not involved in the SED fitting process. The residual 
value of NUV is too large to be displayed properly in the lower panel.
\label{F3}}
\end{figure*}

  The first panel (i.e., the top one) in Figure~\ref{F2} shows the 
power spectral density of the \textit{TESS} light curve in sector 14.  
The light curve in the second panel is folded with the orbital 
period of 0.600303 days (i.e., folded by using the period from 
radial-velocity fitting); the folded light curve shows the ellipsoidal-like 
variations. We interpret it as a result of the tidal deformation of the 
visible star. The two peaks in the folded light curve have different heights, 
which may be caused by the stellar spots. The light curve and 
radial-velocity curve in Figure~\ref{F2} are folded by using the same 
ephemeris, i.e., ${\rm T}(\phi = 0)=2458631.569431\,{\rm HJD} + 0.600303 
{\rm \times N}$, where the phase $\phi = 0$ corresponds to the visible star 
in the superior conjunction, and HJD is the Heliocentric Julian Date. Light 
curves of all other sectors are folded by the same period of 0.6003 days; 
they also show periodic variations (see Section~\ref{sec3.4}).

 \begin{figure*}[htb!]
\plotone{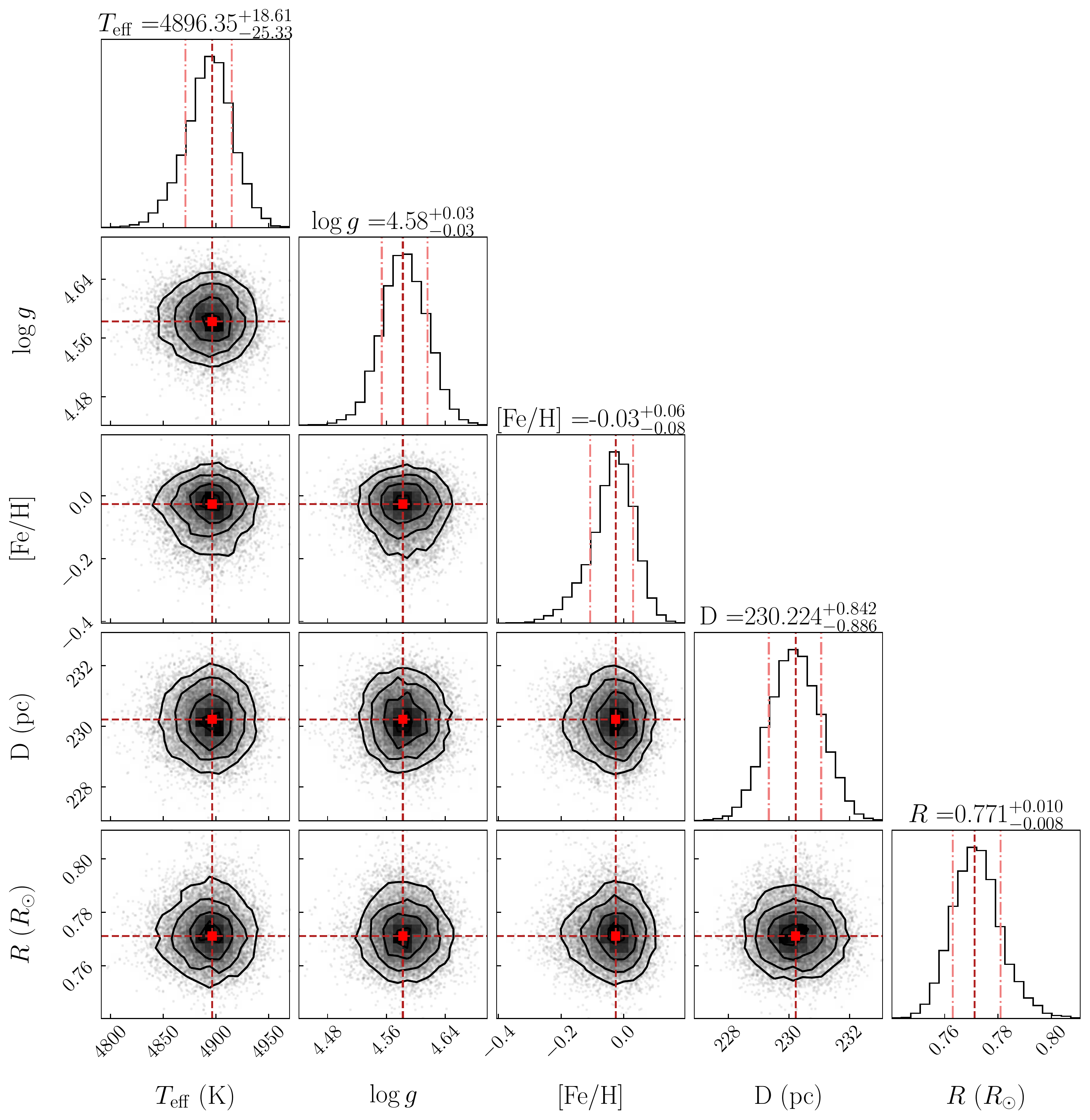}
\caption{The distributions of the SED-fitting parameters. The black dots
represent the nested MCMC samples, and the black curves are the contours
with 1$\sigma$, 2$\sigma$, and 3$\sigma$ probabilities, respectively. The
red points are the best-fitted values.
\label{F4}}
\end{figure*}

\begin{figure*}[htb!]
\plotone{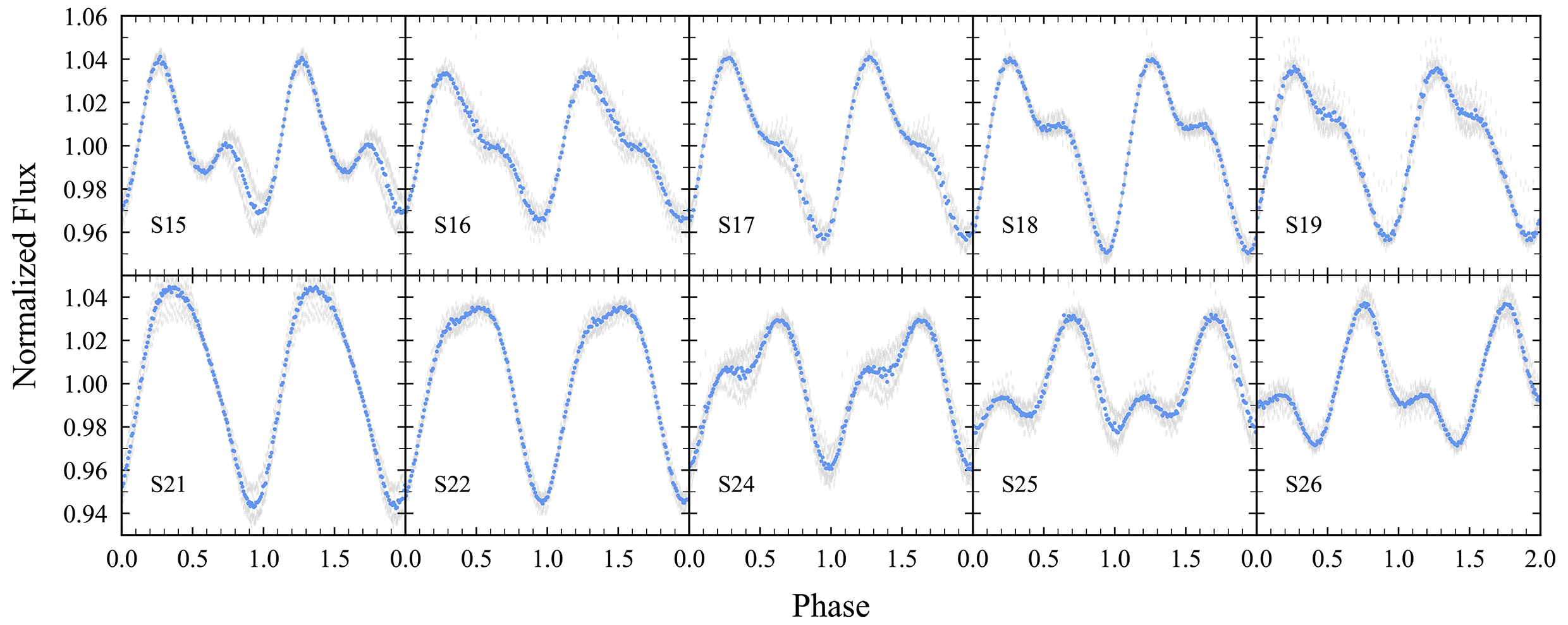}
\caption{\textit{TESS} light curves folded with the orbital period of 0.6003 
days. The light curves in the phases of [1,\,2] are the duplicates of those
[0,\,1] for clarity. The marker in each panel indicates the observed sector
(e.g., S15 represents sector 15). The gray curves show the folded light
curves, and the blue dots are the mean values over ten gray points.
\label{F5}}
\end{figure*}

\section{Results}\label{sec3}
\subsection{Mass function}\label{sec3.1}
  The mass function (the minimum mass of the invisible star) can be
calculated by the equation:
\begin{equation}
\it f(\it M_{\rm{2}}) = \frac{(\it M_{\rm{2}}\sin i)^{\rm 3}}{(\it M_{\rm{1}}+\it M_{\rm{2}})^{\rm 2}} = \frac{\it K^{\rm 3}\it P_{\rm{orb}}(\rm 1-\it e^{\rm 2})^{\rm 3/2}}{\rm 2 \rm \pi \it G},
\label{E4}
\end{equation}
where $M_{\rm{1}}$ and $M_{\rm{2}}$ are the masses of the visible star and
its companion, $i$ is the orbital inclination angle, $e$ is the eccentricity 
(fixed to zero), and $G$ is the gravitational constant. The 
calculated value is $f(M_{\rm{2}}) = 0.120\,\pm\,0.003\,M_{\odot}$.

\subsection{Properties of the visible star}\label{sec3.2}
  We derive the stellar parameters of the visible star by spectral energy
distributing (SED) fitting. The fitting is made by using the Python tool,
ARIADNE \citep[spectrAl eneRgy dIstribution bAyesian moDel averagiNg fittEr;][]{Vines2022}
\footnote{\url{https://github.com/jvines/astroARIADNE}}. We collect the
photometric data in various bands from different sky surveys. The data from
\textit{GALEX} \citep{Martin2005,Morrissey2007}, SDSS \citep{Abazajian2009},
APASS \citep{Henden2009,Henden2010}, Pan-STARRS
\citep{Chambers2016,Magnier2020A,Waters2020}, \textit{Gaia} \citep{Gaia2018},
\textit{TESS} \citep{Stassun2019}, 2MASS \citep{Skrutskie2006}, and ALLWISE
\citep{Cutri2021} are queried via the Vizier
catalog\footnote{\url{https://vizier.u-strasbg.fr/viz-bin/VizieR-4}} with a
maximum matching radius of 2$\arcsec$, and are taken as the input parameters
to ARIADNE. The collected photometric data and the corresponding calculated
flux are listed in Table~\ref{tbl:sed}. The two parameters, the parallax of
4.34 $\pm$ 0.02 mas from \textit{Gaia} DR3 \citep{Gaia2022} 
and the extinction value of $A_{\rm v}$ = 0.09 from \citet{Amores2021} in the
GALExtin\footnote{\url{http://www.galextin.org/}} are fixed in our fit. Note
that we implicitly assume that the observed SED is dominated by a single star
(i.e., the visible star) since it is a single-lined system.

  The SED fitting result is displayed in Figure~\ref{F3}. The green 
curve in Figure~\ref{F3} is obtained by weighting the four stellar atmosphere 
models, which are Phoenix v2\footnote{\url{ftp://phoenix.astro.physik.uni-goettingen.de/HiResFITS/PHOENIX-ACES-AGSS-COND-2011/}}, BT-Models\footnote{\url{http://osubdd.ens-lyon.fr/phoenix/}},
Castelli \& Kurucz\footnote{\url{http://ssb.stsci.edu/cdbs/tarfiles/synphot3.tar.gz}},
and Kurucz 1993\footnote{\url{http://ssb.stsci.edu/cdbs/tarfiles/synphot4.tar.gz}},
and the corresponding weights are 0.24, 0.08, 0.59, and 0.09, as determined 
by the ARIADNE code. The black dots and gray circles in the upper panel each 
represent the observed and synthetic fluxes. The residuals (triangles in the 
lower panel of Figure~\ref{F3}) are defined as ($f_{\rm p} - f_{\rm s}) / f_{\rm s}$, where $f_{\rm p}$ and $f_{\rm s}$ are the photometric and synthetic fluxes, 
respectively. The best-fitting parameters (see Figure~\ref{F4}) and their 
$1\sigma$ uncertainties are obtained from the MCMC chains; the results show 
the spectral type of the visible star is K2V, and the parameters are:
$T_{\rm eff} = 4896^{+19}_{-25}\ \mathrm{K}$, $\log g$ = $4.58 \pm 0.03$ dex,
[Fe/H] = $-0.03^{+0.06}_{-0.08}$, radius
$R_{\mathrm{1}} = 0.77\,\pm\,0.01\,R_{\odot}$. The luminosity
$L_{\rm 1} = 4\,\pi\,R_1^2\,\sigma\,T_{\rm eff}^4 = 0.31^{+0.01}_{-0.02}\,L_{\odot}$.
The radius and luminosity are statistically consistent with those from
\textit{Gaia} DR3: $R_{\rm 1} = 0.810\,\pm\,0.004\,R_{\odot}$ and
$L_{\rm 1} = 0.340\,\pm\,0.006\,L_{\odot}$.

  We calculate the mass of the visible star ($M_{\mathrm{1}}$) by different
methods to check their consistency. First, we take 10,000 samplings
of $\log g$ and the corresponding radius from the ARIADNE's nested
MCMC sampling, and derive the corresponding mass by
$M_{\rm 1} = g R_{\rm 1}^2\,/\,G$. The median value and its uncertainties 
the 16th and 84th percentiles
$M_{\rm 1} = 0.81^{+0.07}_{-0.06}\ M_{\odot}$. Second, we refer to the
mass-luminosity relation of \citet{Henry1993} to calculate the mass
$M_{\mathrm{1}}$ from the infrared J, H, and K luminosity. The corresponding
magnitude of the three bands are 11.516 $\pm$ 0.022 mag, 11.004 $\pm$ 0.021
mag, 10.889 $\pm$ 0.019 mag from 2MASS (Two Micron All Sky
Survey)\footnote{\url{https://old.ipac.caltech.edu/2mass/}}. The derived 
masses of three bands are similar, i.e., 
$M_{\mathrm{1}}^{*} = 0.79\pm 0.1\ M_{\odot}$, which is in accord with the 
mass calculated by the first method.

\subsection{The nature of the invisible object: a compact object or a normal star?}\label{sec3.3}
  The companion in J1729+6529 ought to be a compact object for the following
reasons. We calculate the mass $M_{\rm{2}}$ of the invisible companion by
using the mass function in Equation~(\ref{E4}), where
$M_{\rm 1} = 0.81^{+0.07}_{-0.06}\ M_{\odot}$ from SED and mass function of
$f(M_{\rm{2}}) = 0.120\,\pm\,0.003\,M_{\odot}$. The derived minimum value 
of $M_{\rm{2}}$ is 0.63 $M_{\odot}$, which corresponds to 
$i = 90^{\circ}$. Hence, the mass ratio of J1729+6529 is
$q = M_{\rm{2}} / M_{\rm 1} \gtrsim 0.78$. In such case, the 
contribution of the companion to the optical band should be significant. 
However, CCFs indicate that the spectra of J1729+6529 are single-lined. 
Therefore, we conclude that the invisible companion is a compact object.

  Furthermore, we focus on the absence of occultation in the light curves.
We calculate the binary separation $a$ by
$G\,(M_{1}+M_{2})\,P_{\rm orb}^2 = 4\,\pi^2\,a^3$. We introduce the
distance $\Delta d = R_1 + R_2 - a\cos i$, which is the geometric
difference between the sum of the two stellar sizes and the projection of
the binary separation. If $\Delta d>0$ ($\Delta d<0$), we expect eclipsing
(non-eclipsing) light curves. The critical inclination angle ($i_\mathrm{c}$)
for $\Delta d=0$ depends upon the masses and sizes of two stars and the
period. Meanwhile, the mass of the secondary object can be calculated from
the mass function (see Section~\ref{sec3.1}) and the mass of the primary
star (see Section~\ref{sec3.2}) for each inclination angle. If the secondary
object is a normal star, we can use the mass-radius relation of 
\citet{Eker2018} to derive $R_2$. Hence, we derive a critical inclination 
angle ($i_\mathrm{c}=66^{\circ}$) for $\Delta d=0$. If our binary is a 
late-type stellar companion, we expect that $i\lesssim i_\mathrm{c}$
(i.e., $\Delta d <0$) since the eclipsing signature is absent in the
\textit{TESS} light curve. The corresponding $M_2\gtrsim 0.7\ M_{\odot}$.
In this case, the mass ratio increases to $q \gtrsim 0.86$, which means that 
the contribution of the companion should be even more significant. However,
neither the spectra nor the photometry shows an excess component from the
companion star. This further supports that the invisible companion is a
compact object.

\subsection{Properties of the invisible companion}\label{sec3.4}
  The compact object with the mass of $M_{\rm{2}}\,\gtrsim\,0.63\,M_{\odot}$
in J1729+6529 indicates that it can either be a WD or a NS,
which mainly depends on the inclination $i$. Note that if the compact 
object is a NS with a typical mass 1.28 $\pm$ 0.24$\,M_{\odot}$ 
\citep{Ozel2012}, the inclination angle should be about $39^{+7}_{-5}$ 
degrees according to the mass function. In principle, one can constrain
$i$ by fitting the ellipsoidal variations to the high-cadence \textit{TESS}
light curves. In Figure~\ref{F5}, we plot the \textit{TESS} light curves for
sectors of 15-19, 21-22 and 24-26. It is clear that the light curves of
J1729+6529 show unclear complex features. Possible origin of such complex
features are due to the strong stellar activity. The impact factors on the
light curves are variable, such as stellar surface differential rotation,
rotation period, periodicity of active stars, shear, the number of spots,
spot lifetime, and so on. More details can be referred to  \citet{Basri2011,Basri2018,Basri2020,Walkowicz2013}.

  We argue that the compact object is a WD for the following 
reasons. It is evident that, compared with the best-fitting stellar 
spectral template, the SED has a clear NUV excess, which is well expected 
from a WD. Stellar chromospheric activity might also produce NUV emission 
and its strength is highly correlated with the H$\alpha$ emission 
\citep[e.g.,][]{Jones2016,Linsky2017}. We did not detect the H$\alpha$ 
emission lines in most of the observed spectra of J1729+6529. Therefore, 
we argue that the NUV excess is dominated by the emission of a WD, although 
a small portion might be produced by the chromospheric emission of the visible 
star.

  We constrain the WD's effective temperature by comparing the NUV excess 
with the WD cooling models
\citep{Bergeron1995,Kowalski2006,Bergeron2011,Tremblay2011,Blouin2018,Bedard2020}.
The absolute NUV magnitude of J1729+6529 is 
$M_{\rm NUV}\,=\,m_{\rm NUV}\,-\,5\,\log(D)\,+\,\rm{5}\,-\,\it{A}_{\rm NUV} = \rm {12.15}  ~\mathrm{mag}$, 
where $m_{\rm NUV}$ is the visual NUV magnitude from $GALEX$, $D$ is the 
distance in pc, and $A_{\rm NUV} = 0.25$ is the NUV extinction value derived 
by using the Milky Way Average Extinction Curve \citep{Gordon2009}. The 
synthetic colors for WDs are adopted from \url{http://www.astro.umontreal.ca/~bergeron/CoolingModels}.
We investigate two cooling models of DA-type WDs with typical masses 
$M_{\rm 2} = 0.60\,M_{\odot}$ and $M_{\rm 2} = 0.80\,M_{\odot}$. These two 
masses correspond to orbital solutions with the inclination 
$i \approx 90^{\circ}$ and $i \approx 60^{\circ}$, respectively. As analyzed 
above, the absolute NUV magnitude of the WD should be $\gtrsim$ 12.15 mag. 
Therefore, by looking up the cooling models, only WDs with 
$T_{\rm eff}^{\rm WD} \lesssim$ 12000 K (for $M_{\rm 2}= 0.60\,M_{\odot}$) and 
$T_{\rm eff}^{\rm WD} \lesssim$ 13500 K (for $M_{\rm 2}= 0.80\,M_{\odot}$) can 
satisfy the UV excess. The corresponding WD ages are $\gtrsim\,0.39$ Gyr and 
$\gtrsim\,0.45$ Gyr, respectively.

\section{Conclusions and discussion}\label{sec4}
  J1729+6529 is a single-lined spectroscopic binary selected from the optical
time-domain survey. We summarize our results as follows. First, we obtain the 
radial-velocity semi-amplitude
$K = 124.4^{+1.0}_{-1.1}$ km s$^{-1}$ and orbital period
$P_{\rm orb} = 0.600303(6)$ days from the 
LAMOST and P200 spectra and the \textit{TESS} high-cadence light curves. 
Second, we derive the mass $M_{\mathrm{1}}$ = $0.81^{+0.07}_{-0.06}\,M_{\odot}$
of the visible star via SED fitting. Third, we conclude that the unseen
companion is a compact object owing to $M_{\rm 2} \gtrsim\,0.63\,M_{\odot}$
and the single-lined spectra. Fourth, the compact object is likely to be a
WD ($T_{\rm eff}^{\rm WD}\,\lesssim\,12000-13500$ K) rather than a 
NS unless the inclination angle is small ($i\,\lesssim\,40^{\circ}$). Our 
results suggest that the optical time-domain surveys, such as the joint of 
LAMOST and \textit{TESS}, have great advantages in unveiling the faint WDs 
with bright companions (e.g., A, F, G, and K-type) or those hidden 
in the deep sky, and the method is also applicable to reveal other compact 
objects in binaries such as stellar-mass black holes \citep[e.g.,][]{Yi2019}.

  In addition to optical and UV observations, we also searched for the
possible X-ray counterparts of J1729+6529. The nearest detected X-ray source
(2RXS J172901.3+652948) is in the ROSAT 2RXS Catalog of HEASARC (High 
Energy Astrophysics Science Archive Research Center)
\footnote{\url{https://heasarc.gsfc.nasa.gov/cgi-bin/W3Browse/w3browse.pl}},
whose X-ray flux is ($3.08\,\pm\,0.54$)$\times10^{-13}\,\rm erg\,s^{-1}\,
cm^{-2}$, and the corresponding luminosity is 
$L_{\rm X} = (1.95\,\pm\,0.34)\times10^{30}\,\rm erg\,s^{-1}$. The X-ray 
position is 6.4$\arcsec$ away from the optical position of J1729+6529.
Note that there is another brighter star which is 8.87$\arcsec$ away from
J1729+6529. Considering the angular resolution of ROSAT (is around 
$30\arcsec$), we cannot rule out the possibility that the ROSAT X-ray source 
is related to this nearby bright star. If the X-ray is indeed from 
J1729+6529, it may be emitted from the rapidly rotating K star. The reason 
is that the ratio of X-ray luminosity to the bolometric luminosity 
log ($L_{\rm X}/L_{\rm bol}$) $= -2.79$ is comparable to the rapidly rotating 
K stars \citep[around $-3$; e.g., the sources of Speedy Mic and AB Dor in 
Table 5 of ][]{Singh1999}.

\begin{figure*}[htb!]
\plotone{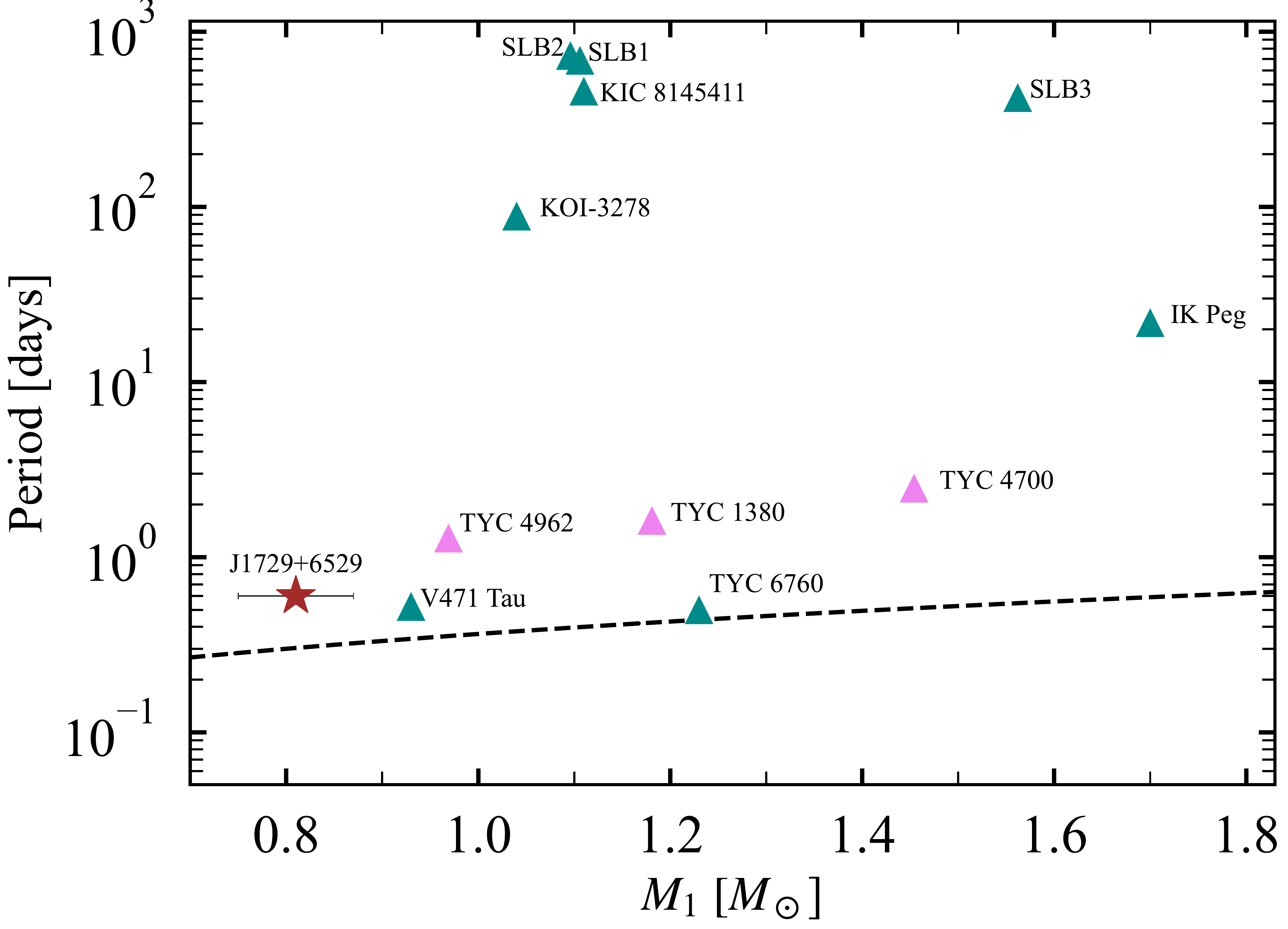}
\caption{The pink triangles represent the sources with detailed 
parameters measured by \citet{Hernandez2021}. The green triangles show the 
A, F, G, K-type WDMS collected by \citet{Hernandez2021}. The red star 
denotes the source J1729+6529. The black dashed line shows the lower limit 
of the orbital period.
\label{F6}}
\end{figure*}

  For comparison, we collected known WDMS systems with A, F, G, K-type 
companions from previous works \citep[e.g.,][note that the cataclysmic variables are not considered for simplicity]{Parsons2016,Rebassa-Mansergas2017,Ren2020}. 
The total number of such systems is around 3000, among which about 1000 systems 
have K-type companions (3800 K $\textless T_{\rm eff} \textless$ 5300 K). 
Most sources in the samples are WDMS candidates 
\citep[except for 9 sources confirmed with HST spectra, ][]{Parsons2016}, which 
were selected by the color index of FUV-NUV and the temperature of the visible 
stars, while the periods of these sources are often unknown. Well-identified 
WDMS binary with a bright companion (A, F, G, and K-type) is rare. We  
found only 11 such sources presented by \citet{Hernandez2021}, of which the 
orbital periods and the masses were robustly measured. The three pink triangles 
in Figure~\ref{F6} represent sources measured by \citet{Hernandez2021}, and 
the green triangles represent other sources collected by 
\citet{Hernandez2021}. By using the mass-radius relation in 
\citet{Eker2018} and the minimum orbital period (Equation (4) in \citealt{Zheng2019}),
we plot a theoretical line of the minimum 
orbital periods $P_{\rm orb}^{\rm min}$ for given masses (black dashed line 
in Figure~\ref{F6}). Figure~\ref{F6} shows that most A, F, G, and K-type 
sources have longer orbital periods compared to J1729+6529 (red star).

We thank the anonymous referee for constructive suggestions that 
improved the paper. This work was supported by the National Key R\&D 
Program of China under grant 2021YFA1600401, and the National Natural Science 
Foundation of China under grants 11925301, 12033006, 11973002, 11988101, 
11933004, 12103041, 12090044, 11833006, 12090042, U1831205, and U1938105. 
This work uses the data collected from multiple telescopes. Guoshoujing
Telescope (the Large Sky Area Multi-Object Fiber Spectroscopic Telescope
LAMOST) is a National Major Scientific Project built by the Chinese
Academy of Sciences. Funding for the project has been provided by
the National Development and Reform Commission. LAMOST is operated
and managed by the National Astronomical Observatories, Chinese Academy
of Sciences. This paper includes data collected by the \textit{TESS}
mission, which is publicly available from the Mikulski Archive for
Space Telescopes (MAST). This research uses data obtained through the
Telescope Access Program (TAP), which has been funded by the TAP member
institutes. This paper also includes public data collected by the NASA
Galaxy Evolution Explorer (\textit{GALEX}), SDSS, the Pan-STARRS1 Surveys
(PS1), the Two Micron All Sky Survey (2MASS), the Wide-field Infrared
Survey Explorer (WISE), the Zwicky Transient Facility (ZTF) project,
ASAS-SN, the European Space Agency (ESA) mission \textit{Gaia}, and the
APASS database.

\begin{table*}
  \footnotesize
  \centering
  \caption{The SED of J1729+6529}
  \label{tbl:sed}
  \begin{tabular}{cccccc}
  \hline
  Telescope  & Band & $\lambda_{\rm central}$  & magnitude & magnitude system & $\lambda f (\lambda)$ \\ \hline
             &   &  \AA &  $ mag $ &  & $\mathrm{\times 10^{-13}erg\ s^{-1}\ cm^{-2}}$ \\ \hline
  (1)       & (2) & (3) & (4) & (5) & (6) \\  \hline
  \multirow{1}{2cm}{GALEX}
&	NUV	&	2313.89 	&	19.211 	$\pm$	0.071 	&	AB	&	9.81 	$\pm$	0.64 	\\
  \hline
  \multirow{1}{2cm}{SDSS}
&	u	&	3561.79 	&	15.719 	$\pm$	0.005 	&	AB	&	153.56 	$\pm$	0.71 	\\
  \hline
  \multirow{4}{2cm}{APASS}
  &	SDSS.g	&	4718.87 	&	13.735 	$\pm$	0.074 	&	AB	&	754.99 	$\pm$	51.46 	\\
&	SDSS.r	&	6185.19 	&	12.866 	$\pm$	0.026 	&	AB	&	1276.00 	$\pm$	30.56 	\\
&	JOHNSON B	&	4347.53 	&	14.223 	$\pm$	0.059 	&	Vega	&	559.36 	$\pm$	30.40 	\\
&	JOHNSON V	&	5504.67 	&	13.168 	$\pm$	0.016 	&	Vega	&	1062.18 	$\pm$	15.65 	\\
  \hline
  \multirow{2}{2cm}{Pan-STARRS}
  &	g	&	4866.46 	&	13.663 	$\pm$	0.013 	&	AB	&	783.63 	$\pm$	9.38 	\\
&	r	&	6214.62 	&	13.078 	$\pm$	0.013 	&	AB	&	1047.36 	$\pm$	12.54 	\\
&	i	&	7544.57 	&	12.793 	$\pm$	0.013 	&	AB	&	1111.71 	$\pm$	13.31 	\\
&	z	&	8679.48 	&	12.642 	$\pm$	0.013 	&	AB	&	1104.24 	$\pm$	13.22 	\\
&	y	&	9633.26 	&	12.506 	$\pm$	0.005 	&	AB	&	1130.46 	$\pm$	5.21 	\\
  \hline
  \multirow{3}{2cm}{Gaia}
  &	BP	&	5128.97 	&	13.464 	$\pm$	0.005 	&	Vega	&	854.84 	$\pm$	3.94 	\\
&	G	&	6424.93 	&	12.944 	$\pm$	0.001 	&	Vega	&	1062.84 	$\pm$	0.98 	\\
&	RP	&	7799.19 	&	12.291 	$\pm$	0.003 	&	Vega	&	1219.66 	$\pm$	3.37 	\\
  \hline
  \multirow{1}{2cm}{TESS}
  &	red	&	7972.36 	&	12.346 	$\pm$	0.007 	&	Vega	&	1231.10 	$\pm$	7.94 	\\
  \hline
  \multirow{3}{2cm}{2MASS}
 &	J	&	12408.38 	&	11.516 	$\pm$	0.022 	&	Vega	&	949.02 	$\pm$	19.23 	\\
&	H	&	16513.66 	&	11.004 	$\pm$	0.021 	&	Vega	&	733.60 	$\pm$	14.19 	\\
&	Ks	&	21655.84 	&	10.889 	$\pm$	0.019 	&	Vega	&	402.03 	$\pm$	7.04 	\\
  \hline
  \multirow{4}{2cm}{ALLWISE}
  &	W1	&	33791.91 	&	10.806 	$\pm$	0.023 	&	Vega	&	132.54 	$\pm$	2.81 	\\
&	W2	&	46292.96 	&	10.816 	$\pm$	0.020 	&	Vega	&	53.05 	$\pm$	0.98 	\\
&	W3	&	123340.00 	&	10.653 	$\pm$	0.046 	&	Vega	&	4.44 	$\pm$	0.19 	\\
&	W4	&	222530.00 	&	$\textgreater$\,9.328			&	Vega	&	$\textless$\,2.06			\\
  \hline
\end{tabular}\\
\textbf{Notes.} The value of central wavelength can be accessed from the filter library of the
PYPHOT\footnote{\url{https://mfouesneau.github.io/docs/pyphot/}}. The magnitude systems are
referred to the Filter Profile Service\footnote{\url{http://svo2.cab.inta-csic.es/theory/fps/}}.
\end{table*}


\begin{thebibliography}{}

\bibitem[Abazajian et al.(2009)]{Abazajian2009} Abazajian, K.~N., Adelman-McCarthy, J.~K., Ag{\"u}eros, M.~A., et al.\ 2009, \apjs, 182, 543. doi:10.1088/0067-0049/182/2/543

\bibitem[Am{\^o}res et al.(2021)]{Amores2021} Am{\^o}res, E.~B., Jesus, R.~M., Moitinho, A., et al.\ 2021, \mnras, 508, 1788. doi:10.1093/mnras/stab2248

\bibitem[Bar et al.(2017)]{Bar2017} Bar, I., Vreeswijk, P., Gal-Yam, A., et al.\ 2017, \apj, 850, 34. doi:10.3847/1538-4357/aa91d4

\bibitem[Basri et al.(2011)]{Basri2011} Basri, G., Walkowicz, L.~M., Batalha, N., et al.\ 2011, \aj, 141, 20. doi:10.1088/0004-6256/141/1/20

\bibitem[Basri \& Nguyen(2018)]{Basri2018} Basri, G. \& Nguyen, H.~T.\ 2018, \apj, 863, 190. doi:10.3847/1538-4357/aad3b6

\bibitem[Basri \& Shah(2020)]{Basri2020} Basri, G. \& Shah, R.\ 2020, \apj, 901, 14. doi:10.3847/1538-4357/abae5d

\bibitem[B{\'e}dard et al.(2020)]{Bedard2020} B{\'e}dard, A., Bergeron, P., Brassard, P., et al.\ 2020, \apj, 901, 93. doi:10.3847/1538-4357/abafbe

\bibitem[Bergeron et al.(1995)]{Bergeron1995} Bergeron, P., Wesemael, F., \& Beauchamp, A.\ 1995, \pasp, 107, 1047. doi:10.1086/133661

\bibitem[Bergeron et al.(2011)]{Bergeron2011} Bergeron, P., Wesemael, F., Dufour, P., et al.\ 2011, \apj, 737, 28. doi:10.1088/0004-637X/737/1/28

\bibitem[Blouin et al.(2018)]{Blouin2018} Blouin, S., Dufour, P., \& Allard, N.~F.\ 2018, \apj, 863, 184. doi:10.3847/1538-4357/aad4a9

\bibitem[Chambers et al.(2016)]{Chambers2016} Chambers, K.~C., Magnier, E.~A., Metcalfe, N., et al.\ 2016, arXiv:1612.05560

\bibitem[Cojocaru et al.(2017)]{Cojocaru2017} Cojocaru, R., Rebassa-Mansergas, A., Torres, S., et al.\ 2017, \mnras, 470, 1442. doi:10.1093/mnras/stx1326

\bibitem[Cui et al.(2012)]{Cui2012} Cui, X.-Q., Zhao, Y.-H., Chu, Y.-Q., et al.\ 2012, Research in Astronomy and Astrophysics, 12, 1197. doi:10.1088/1674-4527/12/9/003

\bibitem[Cutri et al.(2021)]{Cutri2021} Cutri, R.~M., Wright, E.~L., Conrow, T., et al.\ 2021, VizieR Online Data Catalog, II/328

\bibitem[Eker et al.(2018)]{Eker2018} Eker, Z., Bak{\i}{\c{s}}, V., Bilir, S., et al.\ 2018, \mnras, 479, 5491. doi:10.1093/mnras/sty1834

\bibitem[Fontaine et al.(2001)]{Fontaine2001} Fontaine, G., Brassard, P., \& Bergeron, P.\ 2001, \pasp, 113, 409. doi:10.1086/319535

\bibitem[Gaia Collaboration et al.(2018)]{Gaia2018} Gaia Collaboration, Brown, A.~G.~A., Vallenari, A., et al.\ 2018, \aap, 616, A1. doi:10.1051/0004-6361/201833051

\bibitem[Gaia Collaboration(2022)]{Gaia2022} Gaia Collaboration\ 2022, VizieR Online Data Catalog, I/355

\bibitem[Garc{\'\i}a-Berro et al.(1997)]{Garcia1997} Garc{\'\i}a-Berro, E., Ritossa, C., \& Iben, I.\ 1997, \apj, 485, 765. doi:10.1086/304444

\bibitem[Gaskell \& Peterson(1987)]{Gaskell1987} Gaskell, C.~M. \& Peterson, B.~M.\ 1987, \apjs, 65, 1. doi:10.1086/191216

\bibitem[Gordon et al.(2009)]{Gordon2009} Gordon, K.~D., Cartledge, S., \& Clayton, G.~C.\ 2009, \apj, 705, 1320. doi:10.1088/0004-637X/705/2/1320

\bibitem[Gu et al.(2019)]{Gu2019} Gu, W.-M., Mu, H.-J., Fu, J.-B., et al.\ 2019, \apjl, 872, L20. doi:10.3847/2041-8213/ab04f0

\bibitem[Heller et al.(2009)]{Heller2009} Heller, R., Homeier, D., Dreizler, S., et al.\ 2009, \aap, 496, 191. doi:10.1051/0004-6361:200810632

\bibitem[Henden et al.(2009)]{Henden2009} Henden, A.~A., Welch, D.~L., Terrell, D., et al.\ 2009, \aas

\bibitem[Henden et al.(2010)]{Henden2010} Henden, A.~A., Terrell, D., Welch, D., et al.\ 2010, \aas

\bibitem[Henry \& McCarthy(1993)]{Henry1993} Henry, T.~J. \& McCarthy, D.~W.\ 1993, \aj, 106, 773. doi:10.1086/116685

\bibitem[Hernandez et al.(2021)]{Hernandez2021} Hernandez, M.~S., Schreiber, M.~R., Parsons, S.~G., et al.\ 2021, \mnras, 501, 1677. doi:10.1093/mnras/staa3815

\bibitem[Jones \& West(2016)]{Jones2016} Jones, D.~O. \& West, A.~A.\ 2016, \apj, 817, 1. doi:10.3847/0004-637X/817/1/1

\bibitem[Kepler et al.(2019)]{Kepler2019} Kepler, S.~O., Pelisoli, I., Koester, D., et al.\ 2019, \mnras, 486, 2169. doi:10.1093/mnras/stz960

\bibitem[Kilic et al.(2020)]{Kilic2020} Kilic, M., Bergeron, P., Kosakowski, A., et al.\ 2020, \apj, 898, 84. doi:10.3847/1538-4357/ab9b8d

\bibitem[Kowalski \& Saumon(2006)]{Kowalski2006} Kowalski, P.~M. \& Saumon, D.\ 2006, \apjl, 651, L137. doi:10.1086/509723

\bibitem[Li et al.(2014)]{Li2014} Li, L., Zhang, F., Han, Q., et al.\ 2014, \mnras, 445, 1331. doi:10.1093/mnras/stu1798

\bibitem[Li et al.(2021)]{Li2021} Li, C.-. qian ., Shi, J.-. rong ., Yan, H.-. liang ., et al.\ 2021, \apjs, 256, 31. doi:10.3847/1538-4365/ac22a8

\bibitem[Linsky(2017)]{Linsky2017} Linsky, J.~L.\ 2017, \araa, 55, 159. doi:10.1146/annurev-astro-091916-055327

\bibitem[Liu et al.(2012)]{Liu2012} Liu, C., Li, L., Zhang, F., et al.\ 2012, \mnras, 424, 1841. doi:10.1111/j.1365-2966.2012.21285.x

\bibitem[Liu et al.(2020)]{Liu2020} Liu, C., Fu, J., Shi, J., et al.\ 2020, arXiv:2005.07210

\bibitem[Liu et al.(2019)]{Liu2019} Liu, J., Zhang, H., Howard, A.~W., et al.\ 2019, \nat, 575, 618. doi:10.1038/s41586-019-1766-2

\bibitem[Lomb(1976)]{Lomb1976} Lomb, N.~R.\ 1976, \apss, 39, 447. doi:10.1007/BF00648343

\bibitem[Magnier et al.(2020)]{Magnier2020A} Magnier, E.~A., Chambers, K.~C., Flewelling, H.~A., et al.\ 2020, \apjs, 251, 3. doi:10.3847/1538-4365/abb829

\bibitem[Martin et al.(2005)]{Martin2005} Martin, D.~C., Fanson, J., Schiminovich, D., et al.\ 2005, \apjl, 619, L1. doi:10.1086/426387

\bibitem[Maxted et al.(2009)]{Maxted2009} Maxted, P.~F.~L., G{\"a}nsicke, B.~T., Burleigh, M.~R., et al.\ 2009, \mnras, 400, 2012. doi:10.1111/j.1365-2966.2009.15594.x

\bibitem[Merle et al.(2017)]{Merle2017} Merle, T., Van Eck, S., Jorissen, A., et al.\ 2017, \aap, 608, A95. doi:10.1051/0004-6361/201730442

\bibitem[Morrissey et al.(2007)]{Morrissey2007} Morrissey, P., Conrow, T., Barlow, T.~A., et al.\ 2007, \apjs, 173, 682. doi:10.1086/520512

\bibitem[Mu et al.(2022)]{Mu2022} Mu, H.-J., Gu, W.-M., Yi, T., et al.\ 2022, Science China Physics, Mechanics, and Astronomy, 65, 229711. doi:10.1007/s11433-021-1809-8

\bibitem[Oke \& Gunn(1982)]{Oke1982} Oke, J.~B. \& Gunn, J.~E.\ 1982, \pasp, 94, 586. doi:10.1086/131027

\bibitem[{\"O}zel et al.(2012)]{Ozel2012} {\"O}zel, F., Psaltis, D., Narayan, R., et al.\ 2012, \apj, 757, 55. doi:10.1088/0004-637X/757/1/55

\bibitem[Parsons et al.(2016)]{Parsons2016} Parsons, S.~G., Rebassa-Mansergas, A., Schreiber, M.~R., et al.\ 2016, \mnras, 463, 2125. doi:10.1093/mnras/stw2143

\bibitem[Peterson et al.(1998)]{Peterson1998} Peterson, B.~M., Wanders, I., Bertram, R., et al.\ 1998, \apj, 501, 82. doi:10.1086/305813

\bibitem[Price-Whelan et al.(2017)]{Price2017} Price-Whelan, A.~M., Hogg, D.~W., Foreman-Mackey, D., et al.\ 2017, \apj, 837, 20. doi:10.3847/1538-4357/aa5e50

\bibitem[Raddi et al.(2017)]{Raddi2017} Raddi, R., Gentile Fusillo, N.~P., Pala, A.~F., et al.\ 2017, \mnras, 472, 4173. doi:10.1093/mnras/stx2243

\bibitem[Rebassa-Mansergas et al.(2012)]{Rebassa2012} Rebassa-Mansergas, A., Nebot G{\'o}mez-Mor{\'a}n, A., Schreiber, M.~R., et al.\ 2012, \mnras, 419, 806. doi:10.1111/j.1365-2966.2011.19923.x

\bibitem[Rebassa-Mansergas et al.(2016)]{Rebassa-Mansergas2016} Rebassa-Mansergas, A., Ren, J.~J., Parsons, S.~G., et al.\ 2016, \mnras, 458, 3808. doi:10.1093/mnras/stw554

\bibitem[Rebassa-Mansergas et al.(2017)]{Rebassa-Mansergas2017} Rebassa-Mansergas, A., Ren, J.~J., Irawati, P., et al.\ 2017, \mnras, 472, 4193. doi:10.1093/mnras/stx2259

\bibitem[Rebassa-Mansergas et al.(2021)]{Rebassa-Mansergas2021} Rebassa-Mansergas, A., Solano, E., Jim{\'e}nez-Esteban, F.~M., et al.\ 2021, \mnras, 506, 5201. doi:10.1093/mnras/stab2039

\bibitem[Ren et al.(2018)]{Ren2018} Ren, J.-J., Rebassa-Mansergas, A., Parsons, S.~G., et al.\ 2018, \mnras, 477, 4641. doi:10.1093/mnras/sty805

\bibitem[Ren et al.(2020)]{Ren2020} Ren, J.-J., Raddi, R., Rebassa-Mansergas, A., et al.\ 2020, \apj, 905, 38. doi:10.3847/1538-4357/abc017

\bibitem[Ricker et al.(2015)]{Ricker2015} Ricker, G.~R., Winn, J.~N., Vanderspek, R., et al.\ 2015, Journal of Astronomical Telescopes, Instruments, and Systems, 1, 014003. doi:10.1117/1.JATIS.1.1.014003

\bibitem[Scargle(1981)]{Scargle1981} Scargle, J.~D.\ 1981, \apjs, 45, 1. doi:10.1086/190706

\bibitem[Silvestri et al.(2006)]{Silvestri2006} Silvestri, N.~M., Hawley, S.~L., West, A.~A., et al.\ 2006, \aj, 131, 1674. doi:10.1086/499494

\bibitem[Singh et al.(1999)]{Singh1999} Singh, K.~P., Drake, S.~A., Gotthelf, E.~V., et al.\ 1999, \apj, 512, 874. doi:10.1086/306788

\bibitem[Skrutskie et al.(2006)]{Skrutskie2006} Skrutskie, M.~F., Cutri, R.~M., Stiening, R., et al.\ 2006, \aj, 131, 1163. doi:10.1086/498708

\bibitem[Stassun et al.(2018)]{Stassun2018} Stassun, K.~G., Oelkers, R.~J., Pepper, J., et al.\ 2018, \aj, 156, 102. doi:10.3847/1538-3881/aad050

\bibitem[Stassun et al.(2019)]{Stassun2019} Stassun, K.~G., Oelkers, R.~J., Paegert, M., et al.\ 2019, \aj, 158, 138. doi:10.3847/1538-3881/ab3467

\bibitem[Taani(2017)]{Taani2017} Taani, A.\ 2017, arXiv:1702.04419

\bibitem[Ting et al.(2019)]{Ting2019} Ting, Y.-S., Conroy, C., Rix, H.-W., et al.\ 2019, \apj, 879, 69. doi:10.3847/1538-4357/ab2331

\bibitem[Tonry \& Davis(1979)]{Tonry1979} Tonry, J. \& Davis, M.\ 1979, \aj, 84, 1511. doi:10.1086/112569

\bibitem[Tremblay et al.(2011)]{Tremblay2011} Tremblay, P.-E., Bergeron, P., \& Gianninas, A.\ 2011, \apj, 730, 128. doi:10.1088/0004-637X/730/2/128

\bibitem[Vines \& Jenkins(2022)]{Vines2022} Vines, J.~I. \& Jenkins, J.~S.\ 2022, \mnras. doi:10.1093/mnras/stac956

\bibitem[Walkowicz et al.(2013)]{Walkowicz2013} Walkowicz, L.~M., Basri, G., \& Valenti, J.~A.\ 2013, \apjs, 205, 17. doi:10.1088/0067-0049/205/2/17

\bibitem[Waters et al.(2020)]{Waters2020} Waters, C.~Z., Magnier, E.~A., Price, P.~A., et al.\ 2020, \apjs, 251, 4. doi:10.3847/1538-4365/abb82b

\bibitem[Wesselius \& Koester(1978)]{Wesselius1978} Wesselius, P.~R. \& Koester, D.\ 1978, \aap, 70, 745

\bibitem[Wiktorowicz et al.(2020)]{Wiktorowicz2020} Wiktorowicz, G., Lu, Y., Wyrzykowski, {\L}., et al.\ 2020, \apj, 905, 134. doi:10.3847/1538-4357/abc699

\bibitem[Yang et al.(2021)]{Yang2021} Yang, F., Zhang, B., Long, R.~J., et al.\ 2021, \apj, 923, 226. doi:10.3847/1538-4357/ac31b3

\bibitem[Yi et al.(2019)]{Yi2019} Yi, T., Sun, M., \& Gu, W.-M.\ 2019, \apj, 886, 97. doi:10.3847/1538-4357/ab4a75

\bibitem[Zong et al.(2018)]{Zong2018} Zong, W., Fu, J.-N., De Cat, P., et al.\ 2018, \apjs, 238, 30. doi:10.3847/1538-4365/aadf81

\bibitem[Zong et al.(2020)]{Zong2020} Zong, W., Fu, J.-N., De Cat, P., et al.\ 2020, \apjs, 251, 15. doi:10.3847/1538-4365/abbb2d

\bibitem[Zhang et al.(2022)]{Zhang2022} Zhang, Z.-X., Zheng, L.-L., Gu, W.-M., et al.\ 2022, \apj, 933, 193. doi:10.3847/1538-4357/ac75b6

\bibitem[Zhao et al.(2012)]{Zhao2012} Zhao, G., Zhao, Y.-H., Chu, Y.-Q., et al.\ 2012, Research in Astronomy and Astrophysics, 12, 723. doi:10.1088/1674-4527/12/7/002

\bibitem[Zheng et al.(2019)]{Zheng2019} Zheng, L.-L., Gu, W.-M., Yi, T., et al.\ 2019, \aj, 158, 179. doi:10.3847/1538-3881/ab449f

\bibitem[Zverko et al.(2007)]{Zverko2007} Zverko, J., {\v{Z}}i{\v{z}}{\v{n}}ovsk{\'y}, J., Mikul{\'a}{\v{s}}ek, Z., et al.\ 2007, Contributions of the Astronomical Observatory Skalnate Pleso, 37, 49

\end{thebibliography}
\end{document}